\documentclass[12pt]{article}

\font\SC=cmcsc10 scaled 1440

\def\rdots{\mathinner{\mkern1mu\raise1pt\vbox{\kern1pt\hbox{.}}\mkern2mu
   \raise4pt\hbox{.}\mkern2mu\raise7pt\hbox{.}\mkern1mu}}
\newcommand{\be}{\begin{equation}}
\newcommand{\ee}{\end{equation}}
\newcommand{\Z}{{\rm Z\kern-.35em Z}}
\newcommand{\bP}{{\rm I\kern-.15em P}}
\newcommand{\Q}{\kern.3em\rule{.07em}{.65em}\kern-.3em{\rm Q}}
\newcommand{\R}{{\rm I\kern-.15em R}}
\newcommand{\h}{{\rm I\kern-.15em H}}
\newcommand{\C}{\kern.3em\rule{.07em}{.55em}\kern-.3em{\rm C}}
\newcommand{\T}{{\rm T\kern-.35em T}}
\newcommand{\D}{{\kern-.5em /}}

\begin{document}

\openup 1.5\jot

\centerline{\SC{A Diffeomorphism-Invariant Cut-off Regularization of the Determinant}}
\centerline{\SC {of a Scalar Particle in a Euclidean Gravitational Field $^*$}}

\vspace{1in}
\centerline{Paul Federbush}
\centerline{Department of Mathematics}
\centerline{University of Michigan}
\centerline{Ann Arbor, MI 48109-1109}
\centerline{(pfed@math.lsa.umich.edu)}
\vspace{4.0in}

$^*$ This work was supported in part by the National Science Foundation under Grant No. PHY-92-04824 .
\vfill\eject

$$ $$
$$ $$

\centerline{{\bf ABSTRACT}}

\vspace{1in}

\indent

Continuing the thrust of our recent work, but with an important new idea, we find a cut-off regularization of the determinant of a scalar particle in a classical Euclidean gravitational field.  The field is assumed asymptotically flat, and the regularization is diffeomorphism-invariant under coordinate changes that are the identity at infinity.  The scalar field is expanded in term of variables that depend on the gravitational field, with wavelet localization of each variable.  A renormalization group structure is thus automatically present.  A similar construction is carried out for the determinant of a scalar field in a background Yang-Mills field.

\vfill\eject

\section{Introduction.}

\ \ \ \ \
We hope this paper will be the first step towards a new formulation of quantum gravity!

In [1] a Euclidean version of the Yang-Mills field theory was studied.  The field was expressed non-linearly in terms of the basic variables.  Wavelets were used in the definition of these variables.  A nice presentation of the use of wavelets for simple field theories (such as $\phi^4$ theory in two and three dimensions) may be found in [2].  The goal of [1] and the present paper is to extend the ideas of [2] to theories with gauge symmetries, such as Yang-Mills theory and gravity.  It is not necessary to read [1] or [2] to understand the present paper, it is self-contained.  But a consideration of [1] and [2]  will suggest to the reader how the present work might be extended to a treatment of quantum gravity; in the present paper the gravitational field is a classical background field.  Traditional treatments of the scalar field determinant herein studied may be found in [3].  We hope the relationship to the traditional treatments will be further clarified in the future.  

There is just one new idea in this paper beyond those exposed in [1].  It is the use of scattering operators to define the variables (as introduced in Subsection 2.3).  But this idea is enough to allow us to treat the diffeomorphism groups much the same as we previously treated the Yang-Mills gauge group.  In [1] each excitation was associated to a point to which it was it was ``grounded".  But the diffeomorphism group moves points around so that the same idea would be hard to extend to gravity.  The use of scattering operators avoids selection of a point for each excitation.  We apply this new idea to calculate the determinant of a scalar particle in a background field, Yang-Mills in Section 3, and gravity in Section 2.  The program to extend the present work to treat quantum gravity will be certainly very complex, but the treatment of the Yang-Mills theory in [1] provides a rather detailed road map.

In the present work one derives our final formulae ((2.62) and (3.9)) by a number of formal developments.  In (2.62) and (3.9) the matrix elements that appear are each diffeomorphism-invariant, gauge-invariant, respectively.  The cut-off theories are defined by limiting the set of wavelets in the indexing of (2.62) and (3.9).  Restricted to a finite set of wavelets (2.62) and (3.9) are mathematically well-defined invariant expressions.  The final task not here undertaken is to make sense of (2.62) and (3.9) in the limit that all wavelet indices are included (the cut-offs are removed).  We have presented a well-defined, invariant, cut-off theory.  Using wavelets means the cut-off theory is not euclidean-invariant, this invariance must be recovered in the limit.  We sacrificed this for gauge invariance in the cut-off theory.

We close with a number of tentative insights that will guide our future work.

\begin{itemize}
\item[1)]  The treatment of the Yang-Mills theory in this paper using scattering operators, compared to the study in [1] using a gauge transformation to a radial axial gauge, is not as powerful.  We feel for the Yang-Mills case the techniques in [1] are better.
\item[2)] The variables defined by using scattering operators are not as well localized as using the techniques in [1], there are infra-red, or long distance effects.  This is why we prefer the methods of [1] for the Yang-Mills case.  These infra-red problems seem basic in gravity, and we only hope we will learn how to control them.  But we do not know at present whether variables constructed as in the present work for quantum gravity are localized enough to generate a meaningful perturbation theory.  The infra-red problems have their root in the denominators of the scattering operators (see (2.20), (2.21)).  A possible infra-red cut-off procedure may be outlined as follows:  
\begin{description}
\item[a)] Set the space to be flat outside some bounded region.
\item[b)] Require the diffeomorphisms to be the identity in this region.
\item[c)] For expansion functions supported in this flat region (one will have to slightly modify the expansion functions to keep them from spilling into the nonflat region, making them of compact support) let $\Omega$, the scattering operator, acting on them be the identity.
\end{description}
Further ideas are required to fully deal with the infra-red problem.
\item[3)] We do not know if Euclidean gravity makes sense physically.  But it would  be much more difficult to work in Minkowski space.
\item [4)]  The most feared problem with quantum gravity is its nonrenormalizability.  We wish to study whether any progress may be made with this difficulty by letting the coupling constant run as a power

\begin{equation}
 {1 \over G(k)} \stackrel{\sim}{=} {1 \over G_0(k)} + ck^2 + \dots \ .
\end{equation}
This has formally somewhat the same effect as working with a higher derivative gravity theory.  We would like to see if there are any other possibilities here than the usual higher derivative gravity theories (that have undesired spectral properties, ghosts in scattering amplitudes).
\item[5)]  The final formulas for determinants, equations (2.62) and (3.9), may both be easily reexpressed in terms of $S$ matrix elements (for scattering in $4+1$ dimensions).
\end{itemize}

\vfill\eject
\section{Scalar Particle in Gravitational Field.}
\setcounter{equation}{0}

\medskip
\noindent
\underline{2.1.  Preliminary Definitions.}

We work in a four-dimensional Euclidean space (but nothing we do depends crucially on the dimension) where the metric is asymptotically flat.
\begin{equation}
g_{\mu\nu}(x) \longrightarrow \delta_{\mu\nu} \ \ {\rm as} \ \ x \ \ {\rm goes \ to \ infinity.}
\end{equation}
$\phi(x)$ is a real scalar field, whose Lagrangian is
\begin{equation}
{\cal L} = {1 \over 2} g^{1/2} g^{\mu\nu} \phi_{,\mu} \phi_{,\nu} + {1 \over 2} m^2 g^{1/2} \phi^2 .
\end{equation}
The action is given as
\be
S = \int {\cal L} d^4 x .
\ee
The field equation is then
\be
-\Delta\phi + m^2\phi = 0 
\ee
where
\be
\Delta = g^{-{1 \over 2}} \partial_\mu \left( g^{1/2} g^{\mu \nu} \partial_\nu \right).
\ee
The action may also be written as 
\be
{1 \over 2} \ \int g^{1/2}  [-\phi \Delta \phi + m^2 \phi^2 ] d^4x .
\ee
We likewise need $\Delta_0$, the flat Laplacian
\be
\Delta_0 = \delta^{\mu\nu} \partial_\mu \partial_\nu .
\ee
We define
\begin{eqnarray}
H &=& -\Delta + m^2 \\
H_0 &=& -\Delta_0 + m^2 \\
V &=& H - H_0 = \Delta_0 - \Delta .
\end{eqnarray}

\bigskip
\noindent
\underline{2.2.  The Formal Field Theory}.

The formal field theory may originally be presented as follows.  The field $\phi(x)$ is expanded in a weighted wavelet basis:
\be
\phi(x) = \sum \alpha_i u_i(x)
\ee
where we let $\psi_i$ be an orthonormal set of real wavelets on $R^4$ (see [1] for definition of wavelets)
\be 
\left< \psi_i,\psi_j \right> = \int d^4 x \psi_i(x) \psi_j(x) = \delta_{ij} 
\ee
and
\be
u_i(x) = \left( {1 \over {\sqrt{H_0}}} \ \psi_i \right) (x) .
\ee
An expectation of a function of the $\phi's$, {\it p}, is defined as
\be
< {\it p} > = \ {[{\it p}] \over [1]}
\ee
where
\be
[{\it p}] = \prod_i \left( \int^\infty_{-\infty} d\alpha_i \right) e^{-S(\alpha's)} \ {\it p}(\alpha's)N
\ee
([1] is also viewable as $Z$, the partition function.)  $N$ is a functional of the gravitational field added to make the measure
\be
\prod_i \left( \int^\infty_{-\infty} d\alpha_i \right) N
\ee
diffeomorphism-invariant.  We use the fact that the inner product
\be
\int \phi(x) g^{1/2}(x) \phi(x) d^4x
\ee
is diffeomorphism-invariant to deduce that with $N$ given by the highly formal determinant 
\be
N^2 = Det\left( \left< u_i, \ g^{1/2} \; u_ j \right> \right)
\ee
(2.16) becomes formally diffeomorphism-invariant.  We note that our formal construction of the field theory for $\phi$ can be carried out for any basis set, $u_i$, they need not be defined in terms of wavelets.  The somewhat mysterious factor, $N$, here introduced, will be later seen to make a crucial contribution to the diffeomorphism-invariance of the well-defined cut-off theory to be presented.

\bigskip
\noindent
\underline{2.3.  The Change of Variables (Basis).}

We change variables from the $\alpha_i$ to the set $\beta_i$ where
\be
\phi(x) = \sum \alpha_i u_i(x) = \sum \beta_i w_i(x).
\ee
The $w_i(x)$ will be functionals of the gravitational field.  To define these new basis functions we first define some scattering operators.  The familiar scattering operators $\Omega_\pm$ are defined by
\be
\psi_\pm(k) = \Omega_\pm \; e^{ikx} = e^{ikx} + {1 \over k^2 - H_0 \pm i\varepsilon} \ V \psi_\pm(k) 
\ee
or 
\be
\psi_\pm(k) = \Omega_\pm \; e^{ikx} = e^{ikx} + {1 \over k^2 - H \pm i\varepsilon} \ V e^{ikx} .
\ee
These operators describe scattering by the gravitational field of the scalar field $\phi$. (It is really the time independent scattering in a $4+1$ dimensional space, with a fictitious time variable the fifth dimension.)  We instead will need the scattering operator $\Omega$
\be
\Omega = {1 \over 2}(\Omega_+ + \Omega_-).
\ee
More generally we could have used
\be
\Omega_c = c\Omega_+ + \bar{c} \Omega_- \ .
\ee
$\Omega$ has the property that applied to a real function it yields a real function, and this is why we use $\Omega$ instead of $\Omega_+$ or $\Omega_-$.  $\Omega$ is studied further in the next subsection.

$w_i$ is now defined as
\be
w_i(x) = (\Omega \ u_i)(x)
\ee
so if
\be
u_i(x) = {1 \over (2\pi )^2} \int d^4 k \tilde{u}_i(k)e^{ikx}
\ee
then
\be
w_i(x) = {1 \over (2\pi )^2} \int d^4 k \tilde{u}_i(k) {1 \over 2}[\psi_+(k) + \psi_-(k) ].
\ee

To find an expression for the Jacobian of the coordinate change from the $\alpha$'s to the $\beta$'s, we define
\be
v_i(x) \equiv \left( \sqrt{H_0} \psi_i \right) (x) .
\ee
From (2.19) there follows
\be
\alpha_i = \sum_j \beta_j \left< v_i \; , \; w_j \right> .
\ee
The Jacobian is now given as
\be
Det \left( {\partial\alpha_i \over \partial\beta_j} \right) = Det \left(  \left< v_i \; , \; w_j \right> \right) \equiv R.
\ee

\bigskip
\noindent
\underline{2.4.  Diffeomorphisms and Properties of $\Omega$.}

\begin{itemize}
\item[I)] As already pointed out $\Omega$ is a \underline{real operator}, carrying real functions into real functions.
\item[II)]  In the usual potential theory situation $\Omega_+$ and $\Omega_-$ are \underline{partial isometries}.  $\Omega$ would not be a partial isometry in the potential scattering situation.  In the current setting $\Omega_+, \Omega_-$, and $\Omega$ are all not partial isometries since $H$ and $H_0$ are hermitian in different inner products.  $\Omega_+$ and $\Omega_-$ however are partial isometries viewed as mappings from the space with one inner product to the space with the other.)
\item[III)] The $\Omega$'s are all \underline{twisting operators} between $H$ and $H_0$, so that
\be
H\Omega = \Omega H_0
\ee
and generally
\be
f(H)\Omega = \Omega f(H_0).
\ee   
\item[IV)]  We consider \underline{diffeomorphisms} (or coordinate changes) that are the identity at infinity.  We let ${\cal G}$ be a coordinate transformation whose corresponding action on fields is given by $T^{\cal G}$.
\end{itemize}
The action on the gravitational and scalar fields is decribed by
\be
(T^{\cal G} g_{\mu\nu})(x) = {\partial x^{'\alpha} \over \partial x^\mu}
 \ {\partial x^{'\beta} \over \partial x^\nu}\; g_{\alpha \beta}(x')
\ee
\be
(T^{\cal G}\phi)(x) = \phi(x').
\ee
The operators $\Omega$ depend on the gravitational field,$\Omega = \Omega(g_{\mu\nu})$, and we study how these operators are changed by a coordinate transformation.  
\begin{eqnarray}
\left( \Omega \left(T^{\cal G} g_{\mu\nu}\right)f \right)(x) &=& \left( \Omega (g_{\mu\nu}) f \right) (x') \nonumber \\
&=& \left(T^{\cal G} \left(\Omega (g_{\mu\nu}) f \right)\right)(x).
\end{eqnarray}
The operator $H$ of (2.8) has a similar transformation
\be
H(T^{\cal G}g_{\mu\nu}) T^{\cal G} = T^{\cal G}\ H(g_{\mu\nu} ).
\ee
Equation (2.34) follows from the fact that it holds for each of $\Omega_+$ and $\Omega_-$ separately.  We look at this equation for $\Omega_+$
\begin{eqnarray}
\Omega_+ \left( T^{\cal G}g_{\mu\nu}\right)f &=& \lim_{t\rightarrow \infty} e^{-itH(T^{\cal G}g_{\mu\nu})} e^{itH_0}f \\
&=& \lim_{t\rightarrow \infty} e^{-itH(T^{\cal G}g_{\mu\nu})}T^{\cal G} e^{itH_0}f
\end{eqnarray}
This last equation follows since $T^{\cal G}$ is the identity at infinity and $e^{itH_0}f$ lives at infinity as $t \rightarrow \infty$.  (2.37) now becomes
\be
T^{\cal G}\lim_{t\rightarrow \infty} e^{-itH(g_{\mu\nu})} e^{itH_0}f
\ee
using (2.35).  This gives equation (2.34).

One may note the argument we have just given may be replaced by the observation   
that a scattering solution with only outward scattered waves at infinity becomes a scattering solution with only outward scattered waves at infinity under a coordinate transformation that is the identity at infinity.

\bigskip
\noindent
\underline{2.5. The Determinant.}

If we express the action $S$ in terms of the new variables, $\beta_i$, using (2.19), (2.24), (2.6)-(2.9) and (2.31), we get
\be
S = {1 \over 2} \sum \beta_i {\cal M}_{ij} \beta_{j}
\ee
where
\be
{\cal M}_{ij} = \int d^4 x (\Omega \psi_i) \sqrt{g} (\Omega \psi_j)
\ee
We set
\be
{\cal M} = \det({\cal M}_{ij})
\ee
We note the dramatic fact that each element of ${\cal M}_{ij}$ is diffeomorphism-invariant!  Returning to (2.15) with the change of variables from the $\alpha_i$ to $\beta_i$ we see
\be
[1] = Z \sim {1 \over ({\cal M})^{1/2}} \ N R
\ee
$N$ from (2.18) and $R$ from (2.29) .  We use ideas from [1] (equations (4.24) through (4.37)) to find an expression for NR that is rigorously diffeomorphism-invariant.  We first make the replacement
\be
N^2 = Det (<u_i, \sqrt{g} u_j >) \sim Det(< \psi_i, \sqrt{g} \psi_j >)
\ee
The two determinants, both very formal, differ by a numerical factor, due to a change of basis from the $u_i$ to the $\psi_i$, that is indepedent of gravitational field.  We consider $N^2 R^2$
\be
N^2R^2= {\rm Det} (<w_i,v_j>) {\rm Det} (< \psi_i, \sqrt{g} \psi_j>) {\rm Det} (<v_i,w_j>) .
\ee
We define
\be
u^s_i = {1 \over H^{s/2}_0 }\psi_i
\ee
\be
w^s_i = \Omega \; u^s_i
\ee
So
\begin{eqnarray}
u_i &=& u^1_i \\
\psi_i &=& u^0_i \\
w_i &=& w^1_i \\
v_i &=& u^{-1}_i
\end{eqnarray}
We define $R^2(s)$
\be
R^2(s)  = {\rm Det}(\left< w^s_i, u^{-s}_j \right>) {\rm Det}(\left< u^{-s}_i, w^s_j \right>)
\ee
\begin{eqnarray}
N^2 R^2(0) &=& {\rm Det}(\left< w^0_i, \psi_j \right>) {\rm Det} (< \psi_i, \sqrt{g} \psi_j>) {\rm Det}( \left< \psi_i, w^0_j \right>) \\
&=& {\rm Det} (< w^0_i, \sqrt{g} w^0_j>) \\
&=& {\rm Det} ({\cal M}_{ij}) = {\cal M} \nonumber
\end{eqnarray}
from (2.40).  Then
\be
N^2R^2 = N^2R^2(1) = N^2R^2(0)  \cdot {R^2(1) \over R^2(0)} = {\cal M} {R^2(1) \over R^2(0)} .
\ee
At this point we have for $Z$ 
\be
Z = {R(1) \over R(0)} .
\ee
$N$  of course was crucial in getting a diffeomorphism-invariant expression for $NR(0)$.  We now will find an invariant expression for the remaining ratio, ${R(1) \over R(0)}$.  For any matrix $D_{ij}$ (near the identity) we write 
\be
{\rm Det}(D_{ij}) = e^{{\rm Tr}\ell n(D_{ij})} = e^{P(D_{ij})} .
\ee
We then have
\be
{R(1) \over R(0)} = e^{\int^1_0 ds {d \over ds} P(< w^s_i, u^{-s}_j >)} .
\ee
We follow the development in [1].
\be
{d \over  ds} P (<w^s_i, u^{-s}_j >) = {d \over d\ell} P(< w^\ell_i,  u^{-s}_j >)\big|_{\ell =s} + {d \over dr} P(< w^s_i, u^{-r}_j >) \big|_{r=s} .
\ee
We now use ${\rm Det}(AB) = {\rm Det}(A){\rm Det}(B)$ in the form
\begin{eqnarray}
P\left(\left< w^\ell_i , \sqrt{g} w^{-s}_j \right> \right) &=& P\left(\left< w^\ell_i , u^{-s}_j \right> \right) + P\left(\left< u^s_j , \sqrt{g} w^{-s}_j \right> \right) \\
P\left(\left< w^s_i , u^{-r}_j \right> \right) &=& P\left(\left< w^s_i ,  u^{-s}_j \right> \right)
+ P\left(\left< u^s_i ,  u^{-r}_j \right> \right)
\end{eqnarray}
and easily evaluate the derivatives in (2.58) using (2.59)-(2.60) (and neglecting again a gravitational field independent factor)
\be
{d \over  ds} P (<w^s_i, u^{-s}_j >) = {d \over d\ell} P(< w^\ell_i,  \sqrt{g} w^{-s}_j >)\big|_{\ell =s} 
\ee
or
\be
Z = e^{\int^1_0 ds {d \over d\ell} P(< w^\ell_i, \sqrt{g} w^{-s}_j >)|_{\ell =s}} .
\ee
Note that $\left< w^\ell_i, \sqrt{g} w^{-s}_j \right>$  is diffeomorphism-invariant.
\vfill\eject

\section{Scalar Particle in a Yang-Mills Field}
\setcounter{equation}{0}

\ \ \ \ \ We parallel the previous section, for an n-tuple of real scalar fields coupled to a Euclidean classical Yang-Mills field, that is asymptotically zero.  (If a theory is originally presented using complex fields it is to be rewritten using real fields.)  Then $A_\mu$ is taken as real anti-hermitian and satisfies
\be
A_\mu(x) \rightarrow 0 \ \ \ {\rm as} \ \ \ x \rightarrow \infty .
\ee
Gauge transformations will be restricted to satisfy the condition that at infinity they are the identity.
\be
\Delta_0 = \partial_\mu \partial_u
\ee
\be
{\cal L} = {1 \over 2} \sum_a \left[ (\partial_\mu - eA_\mu)\phi^a (\partial_\mu - eA_\mu)\phi^a + m^2 \phi^a \phi^a \right] .
\ee
The equation satisfied by $\phi$ is
\be
m^2 \phi - (\partial_\mu - eA_\mu)(\partial_\mu - eA_\mu )\phi \equiv m^2 \phi - \Delta \phi = 0 .
\ee
Parallel to (2.15) we have
\be
[{\it p}] = \prod_i \left( \int^\infty_{-\infty} d\alpha_i \right) e^{-S(\alpha's)} \ {\it p}(\alpha's).
\ee
There is no factor $N$ here.  Again
\begin{eqnarray}
H &=& -\Delta + m^2 \\
H_0 &=& -\Delta_0 + m^2 \\
V &=& H - H_0 .
\end{eqnarray}
$\Omega$ and the change of basis is exactly as before.  (In this case $\Omega_+$ and $\Omega_-$ are partial isometries, but we never use this fact.)  Exactly parallel to the dervation of (2.62)
\be
Z = e^{\int^1_0 ds {d \over d\ell} P\left(\left< w^{a,\ell}_i,  w^{b,-s}_j \right>\right)\big|_{\ell =s}} .
\ee
Matrix columns and rows are double indexed, the pair $i,a$ labels the wavelet and ``color" of the scalar.

\vfill\eject

\centerline{\underline{REFERENCES}}

\begin{itemize}
\item[[1]] P. Federbush, {\it Prog. of Theor. Phys.} 94 (1995), 1135.

\item[[2]] P. Federbush, {\it Bull. of A.M.S.} 17 (1987), 93.

\item[[3]] N.D. Birrell and P.C.W. Davies, ``Quantum Fields in Curved Space" Cambridge Monographs on Mathematical Physics, Cambridge University Press (1982).
\end{itemize}

\end{document}